\documentclass[aps,prl,nobalancelastpage,twocolumn,superscriptaddress,nolongbibliography,10pt]{revtex4-2}

\usepackage{color,amsthm,amsmath,amsxtra,amsfonts,dsfont,graphicx,bm,amssymb}
\usepackage[colorlinks=true,linkcolor=blue, citecolor=blue, urlcolor=blue, bookmarks]{hyperref}
\usepackage{centernot}
\usepackage[dvipsnames]{xcolor}
\usepackage{tikz}
\usepackage{braket}
\usepackage{subfigure}
\usepackage{bm}
\usepackage{bigints}
\usepackage[normalem]{ulem}
\usepackage{bbold}
\usepackage{mathrsfs}

\newcommand{\be}{\begin{equation}}
\newcommand{\ee}{\end{equation}}
\newcommand{\bea}{\begin{eqnarray}}
\newcommand{\eea}{\end{eqnarray}}
\newcommand{\bse}{\begin{subequations}}
\newcommand{\ese}{\end{subequations}}

\newcommand{\prlsection}[1]{{\em {#1}.---~}}

\newcommand{\includescaledgraphics}[1]{\vcenter{\hbox{\scalebox{0.5}{\includegraphics{#1}}}}}

\begin{document}

\title{Temporal Entanglement, Quasiparticles and the Role of Interactions}

\begin{abstract}
In quantum many-body dynamics admitting a description in terms of \emph{noninteracting} quasiparticles, the Feynman-Vernon influence matrix (IM), encoding the effect of the system on the evolution of its local subsystems, can be analyzed exactly. For discrete dynamics, the temporal entanglement (TE) of the corresponding IM satisfies an area law, suggesting the possibility of an efficient representation of the IM in terms of matrix-product states. A natural question is whether integrable interactions, preserving stable quasiparticles, affect the behavior of the TE. While a simple semiclassical picture suggests a sublinear growth in time, one can wonder whether interactions may lead to violations of the area law. We address this problem by analyzing quantum quenches in a family of discrete integrable dynamics corresponding to the real-time Trotterization of the \emph{interacting} XXZ Heisenberg model. By means of an analytical solution at the dual-unitary point and numerical calculations for generic values of the system parameters, we provide evidence that, away from the noninteracting limit, the TE displays a \emph{logarithmic} growth in time, thus violating the area law. Our findings highlight the non-trivial role of interactions, and raise interesting questions on the possibility to efficiently simulate  the local dynamics of interacting integrable systems.
\end{abstract}

\author{Giacomo Giudice}
\thanks{These authors contributed equally to this work.}
\affiliation{Max-Planck-Institut f\"ur Quantenoptik, Hans-Kopfermann-Stra{\ss}e 1, D-85748 Garching, Germany}
\affiliation{Munich Center for Quantum Science and Technology (MCQST), Schellingstra\ss{}e~4, D-80799 M\"{u}nchen, Germany}
\author{Giuliano Giudici}
\thanks{These authors contributed equally to this work.}
\affiliation{Institute for Theoretical Physics, University of Innsbruck, Innsbruck A-6020, Austria}
\affiliation{Institute for Quantum Optics and Quantum Information,Austrian Academy of Sciences, Innsbruck A-6020, Austria}
\affiliation{Munich Center for Quantum Science and Technology (MCQST), Schellingstra\ss{}e~4, D-80799 M\"{u}nchen, Germany}
\affiliation{Arnold Sommerfeld Center for Theoretical Physics, University of Munich, Theresienstra{\ss}e 37, 80333 M\"{u}nchen, Germany}
\author{Michael Sonner}
\thanks{These authors contributed equally to this work.}
\author{Julian Thoenniss}
\affiliation{Department of Theoretical Physics, University of Geneva, Quai Ernest-Ansermet 30, 1205 Geneva, Switzerland}
\author{Alessio Lerose}
\affiliation{Department of Theoretical Physics, University of Geneva, Quai Ernest-Ansermet 30, 1205 Geneva, Switzerland}
\author{Dmitry A. Abanin}
\affiliation{Department of Theoretical Physics, University of Geneva, Quai Ernest-Ansermet 30, 1205 Geneva, Switzerland}
\author{Lorenzo \surname{Piroli}}
\affiliation{Philippe Meyer Institute, Physics Department, \'Ecole Normale Sup\'erieure (ENS), Universit\'e PSL, 24 rue Lhomond, F-75231 Paris, France}

\maketitle


While computing the exact properties of many-body quantum systems out of equilibrium remains a formidable problem, the past decades have witnessed the development of powerful numerical techniques allowing for accurate approximations. This is especially true in one dimension, where the dynamics can be simulated using algorithms based on matrix-product states (MPSs)~\cite{white2004real,perez2006matrix,daley2004time,vidal2007classical,orus2008infinite,banuls2009matrix,schollwock2011density}. Even in this case, however, the generic linear growth of the entanglement entropy~\cite{calabrese2005evolution,polkovnikov2011colloquium} poses a major obstacle for the MPS representation of the time-evolving state~\cite{schollwock2011density}.

When one is interested in the dynamics of local observables, it is natural to expect that much of the information encoded in the wave function is irrelevant, and that alternative approaches can be devised retaining only the data needed to reconstruct the local physics. A promising idea in this direction was put forward in Ref.~\cite{banuls2009matrix} (see also Refs.~\cite{muller-hermes_tensor_2012,hastings2015connecting,frias2022}), which proposed an MPS algorithm to describe the dynamics induced on local subsystems. Crucially, the efficiency of the method is insensitive to the growth of the standard entanglement entropy. Instead, it is related to the so-called \emph{temporal entanglement} (TE)~\cite{lerose2021influence}, which is naturally understood as the entanglement generated along a space-time rotated direction~\cite{banuls2009matrix}. This approach has recently received renewed interest in connection to the study of space-time dualities in Floquet kicked Ising chains~\cite{akila2016particle,bertini2019entanglement} and dual-unitary quantum circuits~\cite{bertini2019exact}, see Refs.~\cite{piroli2020exact,bertini2020scrambling,rather2020creating,gutkin2020exact,bertini2020operator,bertini2020operator_II,claeys2020maximum,jonay2021triunitary,kos2021correlations,bertini2021random,claeys2021ergodic,reid2021entanglement,prosen2021many,suzuki2021computational}. In addition, similar ideas motivated related constructions exploiting space-time rotation in generic quantum-circuit dynamics~\cite{cotler2018superdensity,lu2021spacetime,garratt2021many,ippoliti2021postselection,zhou2021space,ippoliti2022fractal}.

Recently, the approach developed in Ref.~\cite{banuls2009matrix} has been understood in more physical terms based on the so-called Feynman-Vernon Influence Matrix (IM) approach~\cite{lerose2021influence}, where one views the system as an environment for its local subsystems. Complete information on the local dynamics is encoded in the IM, which can be thought of as a wave function in a multitime Hilbert space. The TE is the bipartite entanglement entropy of the IM.

For time-discrete evolution, it has been argued that the scaling of the TE provides valuable information about the nature of the dynamics~\cite{lerose2021scaling,sonner2021influence,sonner2020characterizing,lerose2022overcoming}, displaying, for instance, a slow growth in many-body localized phases~\cite{sonner2020characterizing}. Still, despite a few interesting examples~\cite{bertini2018exact,piroli2020exact,klobas2021entanglement,klobas2021exact,klobas2021exact_II}, our understanding of the TE scaling remains largely incomplete.

\begin{figure*}
	\includegraphics[width=1\textwidth]{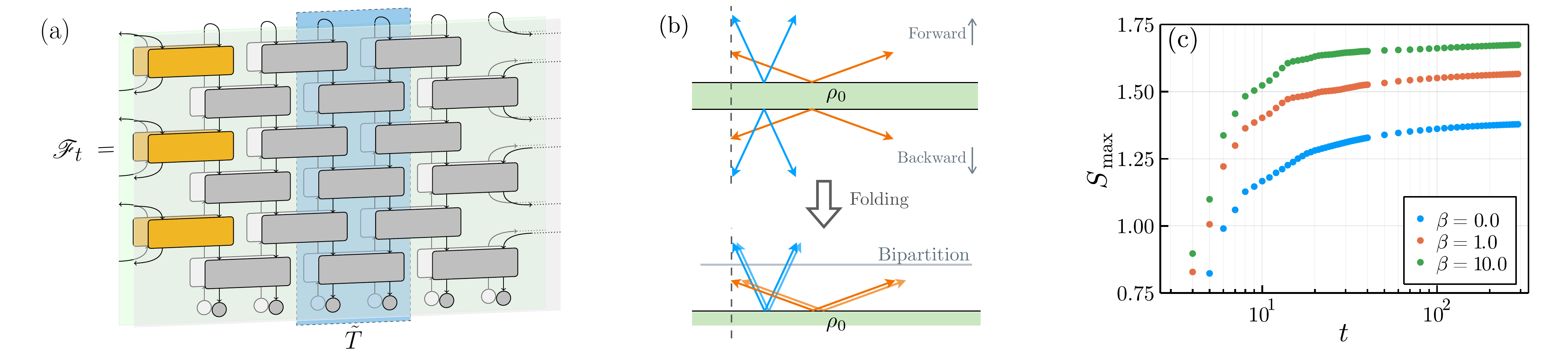}
	\caption{(a) A system of $L$ qubits is initialized in a product state and evolved via a brickwork quantum circuit, with two-site gate defined in Eq.~\eqref{eq:two_site_gate}.  ``Folding'' the circuit, the left and right IMs determine the evolution of local subsystems. Fixing a site $j$, the Floquet operator decomposes into  $\mathcal{U}=\mathcal{U}_{\text{int}} \mathcal{U}_{E}$, cf. the main text. In the figure, $\mathcal{U}_{\text{int}}$ consists of the highlighted gates, while the gray and white gates are part of $\mathcal{U}_{E}$. The operator $\tilde{T}$ defines the dual transfer matrix. (b) Cartoon of the quasiparticle picture for the TE. After folding, backward and forward world lines for each quasiparticle are superimposed, leading to the prediction of sublinear growth for the TE. (c) Growth of the TE for the noninteracting case. In the plot we set $J_x = 0.3, J_y = 0.5, J^\prime=0$, while the initial state is $\rho_0=\bigotimes_k \rho_0^{(k)}$, with $\rho_0^{(k)}=e^{-\beta \sigma^z_k}/\mathcal{Z}$ and $\mathcal{Z}=2\cosh\beta$.}
	\label{fig:general_scheme}
\end{figure*}

As an exception, a detailed characterization of the TE was achieved for \emph{noninteracting} systems, as exemplified for infinite-temperature states in the transverse-field kicked Ising chain~\cite{lerose2021scaling}. Here, the IM was computed analytically, displaying a Bardeen-Cooper-Schrieffer-like structure, and the corresponding TE entropy was shown to obey an \emph{area-law} scaling.

Since the analysis of Ref.~\cite{lerose2021scaling} relies on a quasiparticle description, it is natural to ask about the fate of the TE area law in the presence of \emph{integrable} interactions, preserving the stable quasiparticles. Besides its interest \emph{per se}, this question has implications on the possibility to efficiently simulate the (discrete) dynamics of interacting integrable models, a task known to be hard from the analytical viewpoint~\cite{calabrese2016introduction,essler2016quench,caux2016quench}.

We tackle this question by studying a family of dynamics corresponding to the Trotterization of the \emph{interacting} XXZ Heisenberg model~\cite{vanicat2018integrable,ljubotina2019ballistic}. We focus on quenches from generic initial states, extending the setting of Ref.~\cite{lerose2021scaling} to nonequilibrium situations. Based on a quasiparticle picture~\cite{calabrese2005evolution,fagotti2008evolution,calabrese2009entanglement,calabrese2016quantum,alba2017entanglement,alba2018entanglement}, we argue that the TE scaling is sublinear in time. We provide evidence that, while the area law is preserved for a large class of initial states in the noninteracting case, the TE exhibits a typical \emph{logarithmic} growth in the presence of interactions, violating the area law. We conjecture this to be a generic feature of interacting integrable systems, and discuss some interesting questions raised by our results.

\prlsection{The model} We consider a spin-$1/2$ chain with $L$ sites and periodic boundary conditions. The discrete dynamics is driven by $\mathcal{U}=\mathcal{U}_{\text {odd }} \mathcal{U}_{\text {even }}$, with
\be\label{eq:dynamics}
\mathcal{U}_{\text {odd }}=\prod_{n=1}^{L / 2} U_{2 n, 2 n+1}, \quad \mathcal{U}_{\text {even }}=\prod_{n=1}^{L / 2} U_{2 n-1,2 n}\,,
\ee
where
\be\label{eq:two_site_gate}
U_{n, n+1}=e^{-i {J_x\sigma_{n}^{x} \sigma_{n+1}^{x}-iJ_y\sigma_{n}^{y} \sigma_{n+1}^{y}}-i {J}^{\prime}\left(\sigma_{n}^{z} \sigma_{n+1}^{z}-\openone\right)}
\ee
is a two-site gate expressed in terms of Pauli matrices. We denote by $\ket{0}_j$, $\ket{1}_j$ the states in the local computational basis. Unless stated otherwise, we will set $J_x=J_y=J$. This model was introduced in Refs.~\cite{vanicat2018integrable,ljubotina2019ballistic} as a paradigmatic example of an integrable, periodically driven spin chain and can be thought of as a Trotterized XXZ Heisenberg evolution.
These Floquet dynamics can be represented as a brickwork circuit, cf. Fig.~\ref{fig:general_scheme}(a).

For $J^{\prime}=0$, the model reduces to the XY spin chain, mappable to free-fermion dynamics by a Jordan-Wigner transformation, while for $J_x=J_y=\pi/4$ the circuit generated by repeated application of the Floquet operator $\mathcal{U}$ is dual unitary~\cite{bertini2019exact}. For arbitrary $J$, $J^{\prime}$, the system displays an extensive number of local conservation laws~\cite{vanicat2018integrable,ljubotina2019ballistic} and the Floquet spectrum may be obtained exactly via the Bethe ansatz~\cite{aleiner2021bethe,claeys2021correlations}. The corresponding quasiparticle structure bears similarities to that of the well-known XXZ Heisenberg Hamiltonian~\cite{aleiner2021bethe, korepin1997quantum}.

\prlsection{The quench protocol and the IM} We study a quench, where the system is initialized in product states (either pure or mixed), and analyze the subsequent evolution in the thermodynamic limit $L\to\infty$. The IM formalism~\cite{lerose2021influence} may be introduced starting from the time-evolved expectation value ${\rm Tr}[\rho(t) O_j]={\rm Tr}[\rho_0(\mathcal{U}^{\dagger})^t O_j \mathcal{U}^{t}]$ of a local observable $O_{j}$ at site $j$. Taking for simplicity an initial state $\rho_{0}=\bigotimes_{k} \rho_{0}^{(k)}$, the parts of the system to the left and right of $j$ will be treated as \emph{environments}.
The IMs associated with them arise from integrating out the environment degrees of freedom, treating the trajectory of spin $j$ as an external parameter. Focusing on the right environment $k>j$, we can write down the IM as a Keldysh path integral, where forward and backward spin trajectories are ``folded'' on a closed time contour. We introduce a subsystem-environment decomposition $\mathcal{U}=\mathcal{U}_{\text{int }} \mathcal{U}_{E}$,
where $\mathcal{U}_{\text{int}}$ is the gate acting on spins $j$ and $j+1$, and $\mathcal{U}_{E}$ acts only on spins $k>j$,
which can be done in a natural way exploiting the brickwork structure, cf. Fig.~\ref{fig:general_scheme}(a). Defining the partial matrix elements of $\mathcal{U}_{\text{int}}$ as the operators $[\mathcal{U}_{\text {int}}]_{s,\sigma}=[U_{j,j+1}]^{s}_{\sigma}$ acting on spin $j+1$ only (where $s$, $\sigma$ are the input and output states of spin $j$), the IM $\ket{\mathscr{F}_t}$ is the vector with coordinates
depending on the trajectories $\{s^\pm_\tau,\sigma^\pm_\tau\}$ as
\onecolumngrid
\be\label{eq:IM}
\mathscr{F}_t\left[\sigma_{\tau}^{\pm}, s_{\tau}^{\pm}\right]=\operatorname{Tr}_{E}\left(\left[\mathcal{U}_{\mathrm{int}}\right]_{s_{t}^{+}, \sigma_{t}^{+}} \mathcal{U}_{E} \cdots \mathcal{U}_{E}\left[\mathcal{U}_{\mathrm{int}}\right]_{s_{1}^{+}, \sigma_{1}^{+}} \mathcal{U}_{E} \rho_{0}^{E} \mathcal{U}_{E}^{\dagger}\left[\mathcal{U}_{\mathrm{int}}^{\dagger}\right]_{\sigma_{1}^{-}, s_{1}^{-}} \mathcal{U}_{E}^{\dagger} \cdots \mathcal{U}_{E}^{\dagger}\left[\mathcal{U}_{\mathrm{int}}^{\dagger}\right]_{\sigma_{t}^{-}, s_{t}^{-}}\right)\,,
\ee
\twocolumngrid
\noindent where  $\operatorname{Tr}_{E} \equiv \operatorname{Tr}_{k> j}$ and $\rho_{0}^{E} \equiv \bigotimes_{k> j} \rho_{0}^{(k)}$, cf. Fig.~\ref{fig:general_scheme}.

The IM of a longer environment can be computed from that of a shorter one, leading to an exact self-consistency equation in the thermodynamic limit~\cite{banuls2009matrix,lerose2021influence}. As depicted in Fig.~\ref{fig:general_scheme}, this can be formalized by introducing a dual transfer matrix $\tilde{T}$ generating the evolution in a ``rotated direction'': the self-consistency equation reads $\tilde{T}\ket{\mathscr{F}_t}=\ket{\mathscr{F}_t}$~\cite{banuls2009matrix,muller-hermes_tensor_2012,lerose2021influence} and completely determines $\ket{\mathscr{F}_t}$.

\begin{figure*}
	\centerline{
		\includegraphics[width=\textwidth]{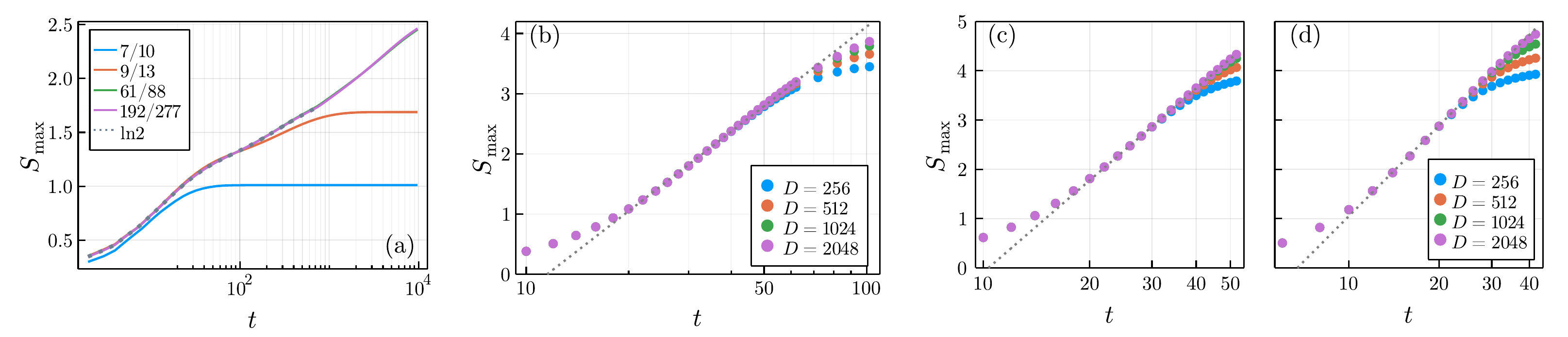}
	}
	\caption{Maximum TE $S_{\rm max}={\rm max}_{\tau}[S_{\tau}(t)]$ as a function of time for different values of $J_x=J_y=\pi/4 + \epsilon$ and $J^\prime = \pi/4 + K$. (a) TE at the dual-unitary point $\epsilon=0$ and $K/\pi=\ln(2)$, quenching from $\ket{\Psi_0}=\ket{+}^{\otimes L}$. The plot is obtained evaluating the entanglement entropy of the analytical MPS solution~\eqref{eq:IM_MPS}. We also show the TE for rational values of $K/\pi$ approximating $\ln(2)$. (b) TE at $\epsilon=0.05$, $J^\prime=1$ for the infinite-temperature state. (c)--(d) Same plot for the N\'eel state $\ket{\Psi_0}=\ket{01}^{\otimes L/2}$. The parameters
	are $J^\prime=1$ and $\epsilon=0.05$ (c), $\epsilon=0.08$ (d). Dotted lines are a guide to the eye to emphasize the logarithmic growth.
	}
	\label{fig:plots_TE}
\end{figure*}

\prlsection{The TE and the quasiparticle picture} The quantity of interest in this work is the TE entropy, denoted by $S_{\tau}(t)$. In order to define it, we consider a bipartition of the multitime Hilbert space of spin trajectories, cut into two regions with time labels $0\le t^\prime \le\tau$ and $\tau+1/2\le t^{\prime\prime} \le t$. Here $t^\prime, t^{\prime\prime} \in (0,t)$ are half integers. The TE is the von Neumann entanglement entropy~\cite{nielsen2002quantum} of the state $\ket{\mathscr{F}_t}$ associated with this bipartition.

We recall that the growth of the standard entanglement entropy after a quantum quench in integrable systems is captured by a well-known quasiparticle picture~\cite{calabrese2005evolution,calabrese2009entanglement,calabrese2016quantum,fagotti2008evolution,alba2017entanglement,alba2018entanglement}. In essence, one postulates that the quench can be modeled as a process creating at each point in space uncorrelated pairs of entangled quasiparticles spreading through the system with opposite momenta. Given two disjoint regions $A$ and $B$, their entanglement then grows proportionally to the number of pairs with one quasiparticle in $A$ and the other in $B$. When supplemented with model-dependent data, this results in a quantitative prediction for the linear growth of the entanglement entropy, which has been proved analytically for noninteracting chains~\cite{fagotti2008evolution} and extensively tested numerically in interacting models~\cite{alba2017entanglement,alba2018entanglement,alba2019entanglement,modak2019correlation,lagnese2021entanglement}. 

Heuristically, we may apply this picture to the TE, cf. Fig.~\ref{fig:general_scheme}(b). Now for each pair we have to keep track of both the forward and backward evolution. Although these trajectories are correlated, they end up being superimposed in the folded spacetime. As a consequence, given a ``space slice,'' all correlated quasiparticles occupy the same temporal position on the Keldysh contour and quasiparticles are not able to transport entanglement at different time sites. One concludes that no TE is generated between disjoint temporal regions after the quench~\footnote{The assumption of the quasiparticles being created in pair is crucial. While this can be justified rigorously for the two-site shift invariant product states considered in this work~\cite{piroli2017what,pozsgay2019integrable}, fine-tuned examples are known for which this is not the case~\cite{bertini2018entanglement,bastianello2018spreading}.}.

A similar heuristic argument already appeared in Ref.~\cite{banuls2009matrix}. However, there it was stated in terms of noninteracting localized excitations and supported by the analysis of a circuit of swap gates~\cite{muller-hermes_tensor_2012}. In contrast, we insist that the picture presented here is in terms of the stable collective quasiparticles of integrable models. As such, it is expected to hold in the scaling limit of large times and to only provide predictions for the leading-order behavior of the TE. That is, the above argument suggests that the TE in integrable systems must asymptotically grow \emph{sublinearly} in time. 

This prediction is consistent with the TE area law scaling found in Ref.~\cite{lerose2021scaling} for the infinite-temperature kicked Ising chain
\be\label{eq:area_law}
{\rm max}_{\tau}[S_{\tau}(t)]\leq c\,,\qquad \forall \, t\,,
\ee
where $c$ is a constant. This result was derived by mapping the system to noninteracting fermions and constructing a gapped quasilocal parent Hamiltonian for $\ket{\mathscr{F}_t}$. The area law~\eqref{eq:area_law} has been numerically confirmed exploiting a covariance-matrix approach for efficient evaluation. A similar analysis can be carried out in our model for $J^{\prime}=0$, for which Eq.~\eqref{eq:dynamics} is mapped onto a free-fermion evolution. In addition, although Eq.~\eqref{eq:area_law} was originally shown for infinite-temperature states~\cite{lerose2021scaling}, the covariance-matrix approach can be generalized to any \emph{Gaussian} initial state~\cite{SM}, allowing us to confirm Eq.~\eqref{eq:area_law} for different values of $J_x$, $J_y$ and various quenches. An example of our data is shown in Fig.~\ref{fig:general_scheme}(c). 

Next, our goal is to test the prediction of the quasiparticle picture and scaling~\eqref{eq:area_law} in the presence of interactions. We provide evidence that, while the TE growth is indeed sublinearly in time, interactions bring about logarithmic violations of the area law.

\prlsection{Exact IM at the dual-unitary point} In principle, integrability allows one to diagonalize the rotated transfer matrix $\tilde{T}$ via the Bethe ansatz and obtain an explicit expression for the IM~\cite{piroli2017quantum,piroli2018non}. However, the resulting wave function is too complicated, and it is not known how to extract the corresponding entanglement.

In order to get some analytical insight, we consider $J=\pi/4$, for which the dynamics is dual unitary~\cite{bertini2019exact}. While in this case the TE is vanishing for a class of fine-tuned initial states~\cite{piroli2020exact}, here we are interested in its behavior for generic ones. To be concrete, we consider a product state $\ket{\Psi_0}=\ket{+}^{\otimes L}$, with $\ket{+}=(\ket{0}+\ket{1})/\sqrt{2}$, although our results generalize to arbitrary two-site shift invariant states $\ket{\Psi_0}=\ket{\psi}_{1,2}\otimes \ket{\psi}_{3,4}\otimes \cdots \otimes \ket{\psi}_{L-1,L}$.

The Bethe ansatz description remains nontrivial at $J=\pi/4$~\cite{aleiner2021bethe}. Nonetheless, the form of the gate in Eq.~\eqref{eq:two_site_gate} becomes simple, allowing us to obtain an exact MPS expression for the IM. Interestingly, we do this avoiding Bethe ansatz techniques and relying instead on methods borrowed from analytical tensor-network theory~\cite{haegeman2017diagonalizing,klobas2021exact}. We consign the details to the Supplemental Material~\cite{SM}, while here we simply report the final result of our analysis. Setting $J^{\prime}=\pi/4+K$, we find that the left IM is
\be\label{eq:IM_MPS}
\bra{\mathscr{F}_{t}}=\braket{v|A^{[1]} B^{[2]} A^{[3]} B^{[4]} \ldots A^{[2t-1]} |w}\,.
\ee
Here, $A$, $B$ are tensors with four physical indices labeled by $00, 01,10,11$ and bond dimension $2t+1$. The corresponding matrices are defined by the elements $[A_{00}]_{\alpha,\beta}=\delta_{\alpha,\beta} \cos[2K\alpha]$, $[A_{01}]_{\alpha,\beta}=\delta_{1,\alpha-\beta} \cos[2K(\alpha-1)]$, $[A_{10}]_{\alpha,\beta}=\delta_{1,\beta-\alpha} \cos[2K\beta]$, $[A_{11}]_{\alpha,\beta}=[A_{00}]_{\alpha,\beta}$
and $[B_{00}]_{\alpha,\beta}=\delta_{\alpha,\beta} \exp[2Ki\alpha]$, $[B_{11}]_{\alpha,\beta}=\delta_{\alpha,\beta} \exp[-2Ki\alpha]$, $[B_{01}]_{\alpha,\beta}=[B_{10}]_{\alpha,\beta}=0$. Here $ \alpha,\beta=-t,-(t-1),\ldots,t$. In addition, the boundary vectors are defined by the elements $\ket{v}_\alpha=\delta_{\alpha,0}$ and $\ket{w}_\alpha=1$. A similar expression holds for the right IM~\cite{SM}.

As an immediate consequence, we obtain
\be\label{eq:upper_bound}
{\rm max}_{\tau}[S_{\tau}(t)]\leq \ln(2t+1)\sim \ln(t)\,,
\ee
yielding a rigorous proof for the sublinear growth of the TE. Here we used that the bipartite entanglement entropy of an MPS with bond dimension $D$ is bounded by $\ln D$~\cite{schollwock2011density}. Despite the simplicity of the solution, the TE displays interesting features. First, we find that the asymptotic behavior at large times is not continuous as a function of $K$. In order to see this, we take $K=\frac{n \pi}{m}$, with $n$, $m$ coprime integers. In this case, it is easy to see that the MPS~\eqref{eq:IM_MPS} can be \emph{compressed} to one with finite bond dimension: because of the periodicity of the trigonometric functions, the infinite matrices $A^{[i]}$ and $B^{[i]}$ can be truncated to the first $m$ lines and columns, so that the TE is bounded. However, this compression is not possible when $K/\pi$ is irrational, suggesting a logarithmic growth.

In order to verify this, we evaluated numerically the TE for irrational values of $K/\pi$, which can be done efficiently since the MPS form of the IM is known exactly. An example of our data is reported in Fig.~\ref{fig:plots_TE}(a), providing evidence of a logarithmic growth. We also show the TE corresponding to rational values approximating $K/\pi$~\footnote{Increasingly better rational approximations are obtained via a continued-fraction expansion  of $K/\pi$}. In other cases, we observe that the TE might display extremely long initial plateaux, which we attribute to the vicinity of $K/\pi$ to rational numbers with small denominator, see Ref.~\cite{SM}. This, in general, makes it challenging to extrapolate the asymptotic behavior from finite-time data. Therefore, while the upper bound~\eqref{eq:upper_bound} is rigorous, given the highly irregular behavior of the TE at the dual unitary point, our numerical evidence should be taken \emph{cum grano salis} in this case. Still, for the accessible timescales, our data consistently point to an indefinite growth of the TE for generic $J$, thus violating the area law~\footnote{Away from the ``singular'' dual-unitary point, we find that the TE behaves in a much more regular way (for instance, no area law appears for rational $K/\pi$), and we expect that the behavior observed within the accessible time scales is representative of the asymptotic one}.


\prlsection{Numerical study for generic interactions} Away from the dual-unitary point, the IM may be obtained using MPS numerical methods. Following Refs.~\cite{banuls2009matrix,muller-hermes_tensor_2012,lerose2021influence}, we represent $\tilde{T}$ as a matrix product operator (MPO), and compute its leading eigenvector using either the density-matrix renormalization group (DMRG)~\cite{schollwock2011density}, or power methods~\cite{banuls2009matrix}. In order to push the available simulation times, we focus on initial states displaying $U(1)$ symmetry, allowing us to enforce it explicitly in the local tensors~\cite{TensorOperations}. Finally, throughout our simulations we used the bond dimension $D$ as a control parameter, checking convergence with respect to it.

We first consider the infinite-temperature state for different values of $J$, $J^{\prime}$. Away from $J=\pi/4$, we find no evidence of a TE area law for rational values of $K/\pi$. In general, we observe an initial linear increase of the TE, followed by an eventual logarithmic growth. Our data are shown in Fig.~\ref{fig:plots_TE}(b): for the available times, curves corresponding to increasing $D$ are seen to converge to a straight line in logarithmic scales.

Next, we turn to the TE from non-equilibrium initial states. In order to preserve $\mathrm{U}(1)$ symmetry, we have chosen the N\'eel state $\ket{\Psi_0}=\ket{01}^{\otimes L/2}$. Here we observe that the TE is large compared to the infinite-temperature state and increasing as we move away from the dual-unitary point. The TE is not symmetric around $t/2$, with its maximum generally attained at later times between $t/2$ and $t$. Its precise location varies with the initial states and parameters. Our numerical data are shown in Fig.~\ref{fig:plots_TE}. Although simulation times are limited, we observe a convincing logarithmic growth emerging after an initial short-time regime. Altogether, our results consistently point to a typical logarithmic violation of the area law in the presence of interactions, which we conjecture to be a general feature of interacting integrable systems. 

\prlsection{Outlook} We have studied the TE in integrable discrete dynamics. Starting from a heuristic quasiparticle picture and based on analytical and numerical evidence in the XXZ Heisenberg model, we have put forward that the TE generically grows logarithmically in time, violating the area law scaling away from the noninteracting regime.

Our findings raise several questions. First, it would be interesting to put our conjecture on rigorous grounds away from the dual-unitary point. From the computational point of view, instead, it would be important to understand whether and how the sublinear growth of the TE may be exploited for an efficient computation of the IM and its approximation in terms of MPS.

A natural question pertains to the relation between the growth of the TE and the operator-space entanglement entropy (OSEE) of local observables~\cite{zanardi2001entanglement,prosen2007operator}. In fact, the latter was also conjectured to grow logarithmically in integrable systems~\cite{prosen2007operator,prosen2007efficiency,pizorn2009operator}; see Refs.~\cite{dubail2017entanglement,alba2019operator,bertini2020operator, bertini2020operator_II,alba2021diffusion} for a proof in special cases. However, at the dual-unitary point of Eq.~\eqref{eq:two_site_gate}, the OSEE was shown to satisfy an area law~\cite{bertini2020operator_II}. Therefore, our results suggest that the OSEE is not directly related to the TE: this is consistent with the intuition that the IM bears information beyond the Heisenberg evolution of local observables, see e.g. Refs.~\cite{klobas2021exact,klobas2021entanglement}.

Finally, while we have focused on discrete dynamics, it would be interesting to study the Trotter limit of continuous-time evolution. Preliminary results suggest that the TE could be \emph{vanishing} in this limit, similarly to the noninteracting case studied in Ref.~\cite{lerose2021scaling}. This would indicate that a continuous MPS ansatz~\cite{verstraete2010continuous,tang2020continuous} could be successfully employed in this limit.

\prlsection{Acknowledgments} L.P. thanks Bruno Bertini for collaboration on closely related projects and many insightful discussions.
G. Giudice and G. Giudici acknowledge support from the Deutsche Forschungsgemeinschaft (DFG, German Research Foundation) under Germany's Excellence Strategy -- EXC-2111 -- 390814868. G. Giudice acknowledges support from the ERC grant QUENOCOBA, ERC-2016-ADG, No. 742102. G. Giudici acknowledges support from the ERC grant QSIMCORR, ERC-2018-COG, No. 771891, and from the Erwin Schr\"odinger Center for Quantum
Science and Technology through a Discovery Grant. This work was supported by the Swiss National Science Foundation (M.S., A.L., D.A.) and by the European Research Council (ERC) under the European Union's Horizon 2020 research and innovation programme (grant agreement No. 864597) (J.T., D.A.). 


\bibliography{bibliography}

\appendix
\clearpage
\newpage

\renewcommand{\theequation}{S\arabic{equation}}
\renewcommand{\thefigure}{S\arabic{figure}}

\onecolumngrid
\begin{center}
	\textbf{\large Supplemental Materials}
\end{center}

Here we provide the technical calculations supporting the results presented in the main text. 

\section{Computation of the TE for free fermions}

In this section, we extend the analysis of Ref.~\cite{lerose2021scaling} to the case of non-equilibrium initial states. We consider the XY-model, defined by Eq.~\eqref{eq:two_site_gate} with $J^\prime=0$. In this case the evolution can be mapped to free fermion dynamics and the IM can be computed exactly. We outline this calculation in the following.\\

As a first step, the spin model is rewritten in terms of fermionic creation and annihilation operators by means of a Jordan-Wigner transformation,
\begin{align*}
\sigma_j^+ &= e^{i\pi \sum_{l<j}c^\dag_l c_l}c_j\qquad
 \sigma_j^- = c_j^\dag e^{-i\pi \sum_{l<j}c^\dag_l c_l}\qquad
\sigma_j^z = (1-2c_j^\dag c_j).
\end{align*}
Making these substitutions in Eq. (\ref{eq:IM}),  one obtains a trace expression over exponentials with quadratic fermionic operators in the exponent. For its evaluation, it is convenient to cast it into a Grassmann path integral such that Gaussian integration techniques can be applied. To this end, we substitute resolutions of the identity in terms of Grassmann coherent states between all operator products in Eq. \eqref{eq:IM}~\cite{lerose2021scaling}. In the environment, we introduce the Grassmann fields $\bar{\bm{\xi}}^\pm_\tau=(\bar{\xi}^\pm_{\tau,n=1},\bar{\xi}^\pm_{\tau,2},..,\bar{\xi}^\pm_{\tau,N})^T$ and $\bm{\xi}^\pm_\tau=(\xi^\pm_{\tau,n=1},\xi^\pm_{\tau,2},..,\xi^\pm_{\tau,N})^T$ at all sites $n$ of the environment at time steps $\tau\in \{0,\tfrac{1}{2},..,t\}.$ The superscript $\pm$ labels the forward and backward branch on the Keldysh contour, respectively. Moreover, we define Grassmann fields $\bar{\eta}^\pm_{\tau}, \eta^\pm_{\tau}$ with $\tau \in \{1,..,t\}$ that analogously describe the degrees of freedom of the system. With this, the IM becomes:

\begin{align}
\nonumber
\mathscr{F}_t\left[\{\bar{\eta}^\pm_\tau,\eta^\pm_\tau\}\right]=&\bigintsss \Big[\prod\limits_{\tau=0,1/2,..}^t d\bar{\bm{\xi}}_\tau^{\pm}d\bm{\xi}_\tau^{\pm}\Big]e^{+\bar{\bm{\xi}}^+_t\bar{\bm{\xi}}^-_t}\Bigg(\bra{\bar{\eta}^+_t,\bar{\bm{\xi}}^+_{t}}\mathcal{U}_{\mathrm{int}}\mathcal{U}_{\text{odd}} \ket{\eta^+_{t},\bm{\xi}^+_{t-1/2}}\cdot e^{-\bar{\bm{\xi}}^+_{t-1/2}\bm{\xi}^+_{t-1/2}}\cdot\bra{\bar{\bm{\xi}}^+_{t-1/2}}\mathcal{U}_\text{even}\ket{\bm{\xi}^+_{t-1}} \\
\nonumber
&\cdots \bra{\bar{\eta}^+_1,\bar{\bm{\xi}}^+_{1}}\mathcal{U}_{\mathrm{int}}\mathcal{U}_{\text{odd}} \ket{\eta^+_{1},\bm{\xi}^+_{1/2}}\cdot e^{-\bar{\bm{\xi}}^+_{1/2}\bm{\xi}^+_{1/2}}\cdot\bra{\bar{\bm{\xi}}^+_{1/2}}\mathcal{U}_\text{even}\ket{\bm{\xi}^+_{0}} e^{-\bar{\bm{\xi}}^+_{0}\bm{\xi}^+_{0}}\cdot \bra{\bar{\bm{\xi}}^+_{0}}\rho_{0}^{E}\ket{\bm{\xi}^-_{0}}\cdot e^{-\bar{\bm{\xi}}^-_{0}\bm{\xi}^-_{0}}\\
\nonumber
&\cdot \bra{\bar{\bm{\xi}}^-_{0}}\mathcal{U}^\dagger_\text{even} \ket{\bm{\xi}^-_{1/2}}\cdot e^{-\bar{\bm{\xi}}^-_{1/2}\bm{\xi}^-_{1/2}}\cdot\bra{\bar{\eta}^-_{1},\bar{\bm{\xi}}^-_{1/2}}\mathcal{U}^\dagger_{\text{odd}}\mathcal{U}^\dagger_{\mathrm{int}} \ket{\eta^-_{1},\bm{\xi}^-_{1}}\\
&\cdots \bra{\bar{\bm{\xi}}^-_{t-1}}\mathcal{U}^\dagger_\text{even} \ket{\bm{\xi}^-_{t-1/2}}\cdot e^{-\bar{\bm{\xi}}^-_{t-1/2}\bm{\xi}^-_{t-1/2}}\cdot\bra{\bar{\eta}^-_{t},\bar{\bm{\xi}}^-_{t-1/2}}\mathcal{U}^\dagger_{\text{odd}}\mathcal{U}^\dagger_{\mathrm{int}} \ket{\eta^-_{t},\bm{\xi}^-_{t}}\Bigg).\label{eq:GM_integral}
\end{align}

To evaluate the single terms in Eq. (\ref{eq:GM_integral}), we first note that $$\bra{\bar{\eta},\bar{\bm{\xi}}}\mathcal{U}_{\mathrm{int}}\mathcal{U}_{\text{odd}} \ket{\eta,\bm{\xi}} = \bra{\bar{\eta},\bar{\xi}_{n=1}}\mathcal{U}_{\mathrm{int}}\ket{\eta,\xi_{n=1}}\cdot \bra{\bar{\tilde{\bm{\xi}}}}\mathcal{U}_{\text{odd}} \ket{\tilde{\bm{\xi}}},$$ where $\bar{\tilde{\bm{\xi}}}$ and $\tilde{\bm{\xi}}$ contain only Grassmann variables for $n>1.$ Next, the Grassmann Kernel of the two-site gates need to be evaluated. For the XY-model with $U_{n,n+1}=\exp\big( -iJ_x \sigma_n^x \sigma_{n+1}^x - i J_y \sigma_n^y \sigma_{n+1}^y\big),$ they read:
\begin{align*}
\nonumber
\mathscr{F}(\bar{\xi}_n,\bar{\xi}_{n+1},\xi_n,\xi_{n+1}) =& \bra{\bar{\xi}_{n},\bar{\xi}_{n+1}} U_{n,n+1}\ket{\xi_{n},\xi_{n+1}}\\
= &\cos (J_x-J_y) \exp\Big[- i \tfrac{\sin (J_x+J_y)}{\cos (J_x -J_y)}  (\bar{\xi}_n \xi_{n+1}+\bar{\xi}_{n+1}\xi_n)+i \tan (J_x-J_y) (\bar{\xi}_{n+1}\bar{\xi}_n + \xi_n \xi_{n+1}) \\
&- 2\tfrac{\sin J_x \sin J_y}{\cos (J_x-J_y)} (\bar{\xi}_n\xi_n+\bar{\xi}_{n+1}\xi_{n+1})\Big] e^{\bar{\xi}_n\xi_n+ \bar{\xi}_{n+1}\xi_{n+1}},
\end{align*}
\twocolumngrid
 where we have suppressed the labels for time index and Keldysh branch.
Furthermore, for a thermal initial state of the form $\rho_0^{E}=\mathop{\otimes}\limits_{n=1}^N e^{-\beta \sigma^z_n}$ (we neglect normalization here since it is not relevant for our purpose), we find for the Grassmann Kernel of the initial state: $\bra{\bar{\bm{\xi}}}\rho_{0}^{E}\ket{\bm{\xi}} = e^{-\beta}\exp\Big( e^{2\beta}\sum\limits_{n=1}^N\bar{\xi}_n\xi_n\Big).$
Making the appropriate substitutions in Eq.  (\ref{eq:GM_integral}), one arrives at an integral of the form:
\begin{align}
\nonumber
&\mathscr{F}_t(\{ \bar{\eta}_\tau, \eta_\tau\}) \sim \bigintsss \Big[\prod\limits_{\tau=0,1/2,..}^t d\bar{\bm{\xi}}^\prime_\tau d\bm{\xi}^\prime_\tau\Big]\\
\nonumber
\Bigg( &\exp\Big[\frac{1}{2}\sum_{\tau,\tau^\prime=0,1/2,..}^t \begin{pmatrix}\bar{\bm{\xi}}^\prime_\tau\\ \bm{\xi}^\prime_\tau\end{pmatrix} \bm{A}_E^{(\tau,\tau^\prime)}  \begin{pmatrix}\bar{\bm{\xi}}^\prime_{\tau^\prime}\\ \bm{\xi}^\prime_{\tau^\prime}\end{pmatrix}\Big]\\
\nonumber
&  \times \exp\Big[\sum_{\tau,\tau^\prime=1}^t \begin{pmatrix}\bar{\bm{\eta}}_\tau\\ \bm{\eta}_\tau\end{pmatrix} \bm{A}_S^{(\tau,\tau^\prime)}  \begin{pmatrix}\bar{\bm{\eta}}_{\tau^\prime}\\ \bm{\eta}_{\tau^\prime}\end{pmatrix}\Big]\\
& \times \exp\Big[\sum_{\tau=1}^t\sum_{\tau^\prime=0,1/2,..}^t \begin{pmatrix}\bar{\bm{\eta}}_{\tau}\\ \bm{\eta}_{\tau}\end{pmatrix} \bm{A}_{\text{int}}^{(\tau,\tau^\prime)} \begin{pmatrix}\bar{\bm{\xi}}^\prime_{\tau^\prime}\\ \bm{\xi}^\prime_{\tau^\prime}\end{pmatrix}\Big]\Bigg),\label{eq:GM_integral_shape}
\end{align} where we defined $\bm{\xi}^\prime_\tau\equiv  (\bm{\xi}^+_\tau,\bm{\xi}^-_\tau)^T$ and $\bm{\eta}_\tau\equiv  (\eta^+_\tau, \eta^-_\tau)^T$ (and analogously for the variables with a bar).
Using standard relations for Gaussian Grassmann integrals, Eq. (\ref{eq:GM_integral_shape}) can be rewritten as:
\begin{align}
\label{eq:IM_free_fermions}
\mathscr{F}_t(\{ \bar{\eta}_\tau, \eta_\tau\}) &\sim \exp\Bigg[\sum_{\tau,\tau^\prime=0}^t \begin{pmatrix}\bar{\bm{\eta}}_{\tau}\\ \bm{\eta}_{\tau}\end{pmatrix}\bm{B}^{(\tau,\tau^\prime)}  \begin{pmatrix}\bar{\bm{\eta}}_{\tau^\prime}\\ \bm{\eta}_{\tau^\prime}\end{pmatrix}\Bigg],
\end{align} with
$$ \mathbf{B}^{(\tau,\tau^\prime)}  = \mathbf{A}_S^{(\tau,\tau^\prime)} + \frac{1}{2}\Big( \mathbf{A}_\text{int}\mathbf{A}_E^{-1}\mathbf{A}_\text{int}^T\Big)^{(\tau,\tau^\prime)}.$$
The state in Eq. (\ref{eq:IM_free_fermions}) can be represented as BCS wavefunction of the form $ \ket{\mathscr{F}_t} \propto \exp(\sum_{i,j} B_{ij}\hat{c}_i^\dagger \hat{c}_j^\dagger) \ket{\emptyset}$, from which we can infer the correlation matrix $\Lambda$ that uniquely determines the IM.  The correlation matrix $\Lambda$ depends only on the matrix $\mathbf{B}$ and is independent of the normalization of  $\ket{\mathscr{F}_t}$ in particular. By defining a bipartition of the system in temporal direction and viewing one part as subsystem, one can compute the TE from the reduced correlation matrix of that subsystem~\cite{latorre2004ground}.  In Fig.~\ref{fig:general_scheme}(c), we show the {\sl maximal} TE, i.e.  for each physical evolution time $t$, we choose the temporal cut in such a way that the TE is maximized.

\section{Exact IM at the dual unitary point}
Here we outline the calculations leading to the exact MPS representation for the IM at the dual-unitary point. 
We consider the following parametrization for the two-site gate
\be
U_{1,2}[K]=e^{-i\left[\frac{\pi}{4} \sigma_1^{x} \sigma_2^{x}+\frac{\pi}{4} \sigma_1^{y} \sigma_2^{y}+\left(\frac{\pi}{4}+K\right)\sigma^{z}_1 \sigma_2^{z}\right]}\,.
\ee
This can be rewritten as
\be
U_{1,2}[K]=S \exp \left[-iK \sigma^{z}_1 \sigma_2^{z}\right]=\exp \left[-iK\sigma^{z}_1 \sigma^{z}_2\right]S\,,
\ee
where $S$ is the swap operator. We focus on quantum quenches from two-site shift invariant product states  $\ket{\Psi_0}=\ket{\psi}_{1,2}\otimes \ket{\psi}_{3,4}\otimes \cdots \otimes \ket{\psi}_{L-1,L}$. Although the method presented in the following can be extended to generic $\ket{\psi}_{j,j+1}$, here we consider the simplest case where $\ket{\Psi_0}$ is one-site shift invariant, and choose for concreteness $\ket{\Psi_0}=\ket{+}^{\otimes L}$, where $\ket{+}=(\ket{0}+\ket{1})/\sqrt{2}$.

We start by analyzing the folded circuit, which we represent pictorially (for $t=2$) as
\be\label{eq:folded_circuit}
\includescaledgraphics{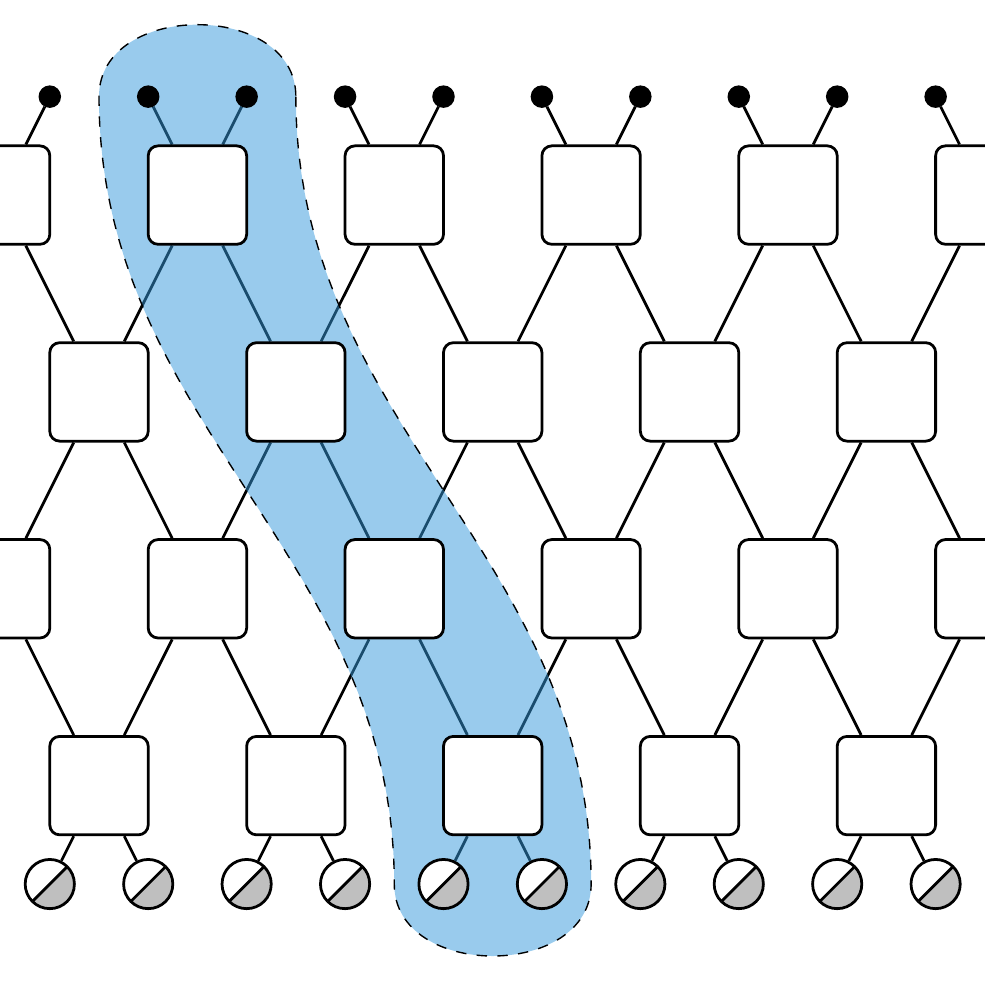}
\ee
where black dots denote the folded identity, while two-tone circles correspond to the folded product state $\ket{+}\ket{+}$. Before tackling the computation of the IM, it is convenient to analyze the so-called light-cone transfer matrix, indicated in Eq.~\eqref{eq:folded_circuit} by a shaded area. It was introduced in Ref.~\cite{gopalakrishnan2019unitary} and it is easy to see that its right fixed point is always a product state of maximally entangled Bell pairs.

In order to compute the left fixed point, we will make use of a graphical ``zipper-equation'', which appeared in Ref.~\cite{haegeman2017diagonalizing} in the context of the classical asymmetric exclusion process. There, it was presented as a tensor-network reformulation of the solution found in Ref.~\cite{derrida1993exact}. In the literature of quantum quenches, zipper equations of similar form were previously exploited to obtain analytical results in interacting quantum cellular automata~\cite{klobas2021entanglement,klobas2021exact,klobas2021exact_II}.

As a starting point, we first look for a solution of the fixed-point equation for a formally infinite transfer matrix. Supposing that the time direction has no boundaries, we may interpret the transfer matrix as obtained by sequential application of the dual gates as
\be
\includescaledgraphics{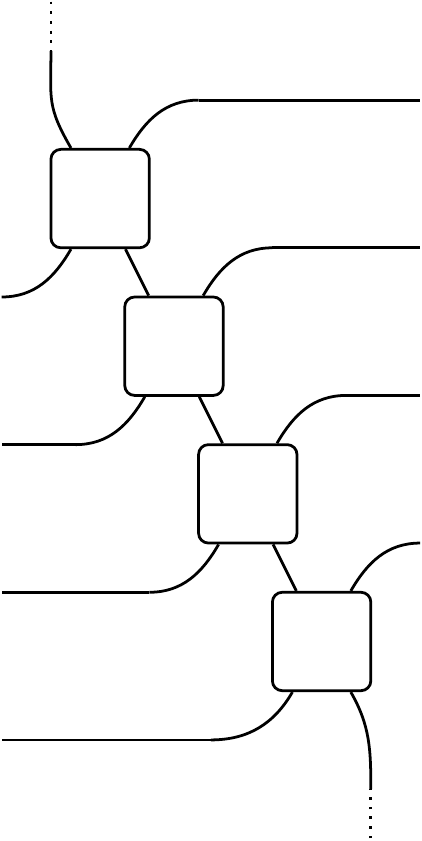}\,.
\ee
We assume that in the bulk the fixed point can be written as an MPS with the same tensor $A$ at each site (its graphical form is given in Eq.~\eqref{eq:MPS_form_LC} below). Following~\cite{haegeman2017diagonalizing}, we observe that a sufficient condition to find a solution is that there exists a tensor $B$ which satisfies the zipper equation
\be\label{eq:intertwinings}
\includescaledgraphics{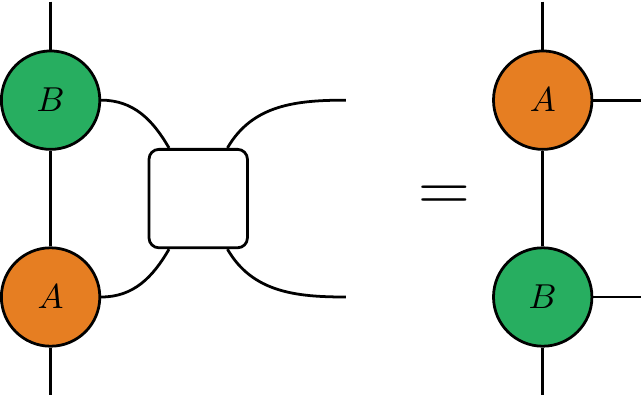}\,.
\ee

Of course, for a finite time $t$, the light-cone transfer matrix has boundaries, which have to be taken into account. Denoting by $\bra{v}$ and $\ket{w}$ the boundary vectors of the MPS, one can see that the following give a sufficient condition for the MPS to be a left fixed point
\begin{subequations}
\label{eq:boundary}
\begin{align}
\includescaledgraphics{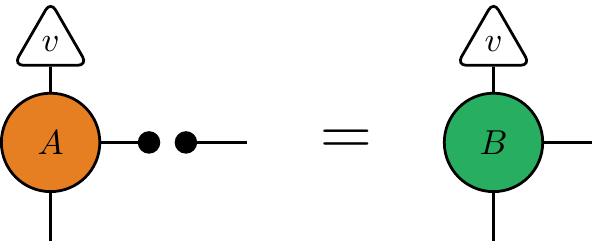}\,, \label{eq:top}\\ \includescaledgraphics{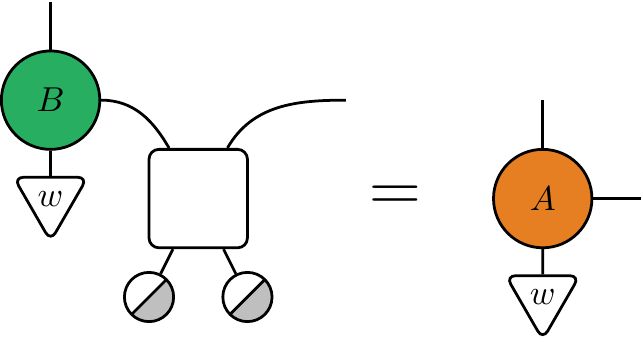}\,.\label{eq:bottom}
\end{align}
\end{subequations}
In the following, we use $k$ to label the four local basis states in the folded picture, i.e. $k=00, 01, 10, 11$. We will find a solution to the bulk and boundary equations~\eqref{eq:intertwinings} and \eqref{eq:boundary}, automatically yielding an MPS solution for the left fixed point.

We note that the first of the boundary conditions~\eqref{eq:boundary} does not depend on the initial state. It states that $\bra{v}B_k=0$ for $k\neq 00,11$. We thus set $B_{01}= B_{10}\equiv 0$, for which it is automatically satisfied. Next, we introduce
\begin{align*}
f(k,\ell)=
\begin{cases}
	0 & {\rm if\ } k,\ell \in \{00,11\}\ \\
	& {\rm or }\ k,\ell \in \{01,10\}\\
	(-1)^{s(k)+s(\ell)} & \mbox{otherwise}\,,
\end{cases}
\end{align*}
where $s(00)=s(10)=1$, $s(01)=s(11)=0$. Eq.~\eqref{eq:intertwinings} can be rewritten as
\be\label{eq:intertwining}
A_{k}B_{\ell}=B_{\ell}A_{k}e^{2iKf(k,\ell)}\,.
\ee
Let $\lambda_0$ be an eigenvalue of $B_{00}$ with eigenvector $\ket{\lambda_0}$. Assuming $\lambda_0\neq 0$, $A_{00}^r\ket{\lambda_0}\neq 0$, we have that $A_{\alpha}^r \ket{\lambda_0}$ is an eigenstate of $B_{00}$ with eigenvalue $e^{-2iKr}$. Excluding the special case where $K/\pi$ is a rational number, the eigenvalues $e^{-2iKr}$ are all different, implying that $B_{00}$ must be infinite dimensional. We thus make the ansatz
\begin{align}
	B_{00}&={\rm diag}(\ldots , e^{-4iK}, e^{-2iK}, 1,e^{2iK}, e^{4iK},\ldots  )\,.
\end{align}
Now, $A_{00}$ and $A_{11}$ commute with $B_{00}$, and since all its eigenvalues are different, they must be diagonal in the same basis, namely
\begin{subequations}\label{eq:a_matrix_01}
\begin{align}
A_{00}&={\rm diag}(\ldots , a_{-1}, a_0, a_1\ldots  )\,,\\
A_{11}&={\rm diag}(\ldots , a^{\prime}_{-1}, a^{\prime}_0, a^{\prime}_1\ldots )\,.
\end{align}
\end{subequations}
On the other hand, $A_{00}$ and $A_{11}$ also commute with $B_{11}$, which we can thus take diagonal in the same basis
\begin{align}\label{eq:b_1}
	B_{11}&={\rm diag}(\ldots , b_{-1}, b_0, b_1\ldots)\,,
\end{align}
where $b_k$ must be determined. Next, using that $A_{01}$, and  $A_{10}$ permute the eigenvectors of $B_{00}$ cyclically [which follows from~\eqref{eq:intertwining}] they must take the form
\begin{subequations}\label{eq:a_matrix}
\begin{align}
	[A_{01}]_{\alpha,\beta}=\delta_{1,\alpha-\beta}\tilde{a}_{\beta}\,, \qquad \alpha,\beta\in \mathbb{Z}\,,\\
	[A_{10}]_{\alpha,\beta}=\delta_{1,\beta-\alpha}\tilde{a}^\prime_{\alpha}\,,\qquad \alpha,\beta\in \mathbb{Z}\,.
\end{align}
\end{subequations}
Eqs.~\eqref{eq:a_matrix}, together with the commutation relations~\eqref{eq:intertwining}, allow us to fix the coefficients $b_\alpha$ in Eq.~\eqref{eq:b_1}, yielding
\begin{align}
	B_{11}&=b_0{\rm diag}(\ldots , e^{4iK}, e^{2iK}, 1,e^{-2iK}, e^{-4iK},\ldots)\,.
\end{align}
where $b_0$ is an overall constant. As it will be manifest later on, for the initial state chosen we can set $b_0=1$.

It remains to fix the constants in Eqs.~\eqref{eq:a_matrix_01}, ~\eqref{eq:a_matrix} and the vectors $\bra{v}$, $\ket{w}$. Using Eq.~\eqref{eq:bottom} for $k=00,11$, we obtain
\be\label{eq:final_1}
[A_{00}]_{\alpha,\beta}=[A_{11}]_{\alpha,\beta}=\frac{1}{2} \cos(2K\alpha)\delta_{\alpha,\beta}\,.
\ee
Next, using  Eq.~\eqref{eq:top}, we get
\be
\ket{v}=(\ldots, 0,0, 1,0,0\ldots)\,.
\ee
Now, Eq.~\eqref{eq:bottom} for $k=10$ read
\be
\frac{1}{4}(B_{00}e^{2iK}+B_{11}e^{-2iK})\ket{w}=A_{10}\ket{w}\,.
\ee
Using~\eqref{eq:a_matrix} and after a little guess work, we immediate see that a solution is given by
\be
\ket{w}=(\ldots, 1,1,1,\ldots)\,,
\ee
and
\be
[A_{10}]_{\alpha,\beta}=\frac{1}{2}\delta_{1,\beta-\alpha} \cos[2K\beta]\,.
\ee
Similarly, repeating the same  steps for $k=01$, we have
\be
\frac{1}{4}(B_{00}e^{-2iK}+B_{11}e^{2iK})\ket{w}=A_{01}\ket{w}\,,
\ee
which straightforwardly yields
\be
[A_{01}]_{\alpha,\beta}=\frac{1}{2}\delta_{1,\alpha-\beta} \cos[2K(\alpha-1)]\,,
\ee
completely fixing the tensors of left fixed point.

Although the MPS solution which we have found is defined in terms of infinite-dimensional matrices, it is immediate to see that the boundary conditions allow one to truncate them at each finite time $t$. Putting all together, for a given time $t$, we have that the left fixed point of the light-cone transfer matrix is given by the MPS
\be\label{eq:MPS_form_LC}
\bra{L_{\rm LC}}=\includescaledgraphics{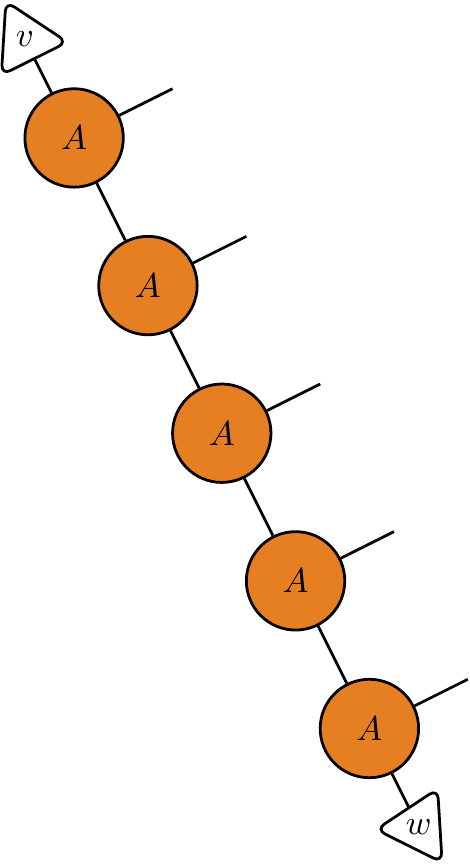}=\braket{v| \underbrace{A A \ldots A}_{2t-1} |w}\,,
\ee
where $v_\alpha=\delta_{\alpha,0}$, $w_\alpha=1$ and (dropping an overall factor $1/2$)
\begin{subequations}
\begin{align}
	[A_{00}]_{\alpha,\beta}&=\delta_{\alpha,\beta} \cos[2K\alpha]\,, \\
	[A_{01}]_{\alpha,\beta}&=\delta_{1,\alpha-\beta} \cos[2K(\alpha-1)]\,,\\
	[A_{10}]_{\alpha,\beta}&=\delta_{1,\beta-\alpha} \cos[2K\beta]\,, \\
	[A_{11}]_{\alpha,\beta}&=[A_{00}]_{\alpha,\beta}\,,
\end{align}
\end{subequations}
with $ \alpha,\beta=-(2t-1), -(2t-2),\ldots ,2t-2 ,2t-1$.

We will use the above result to obtain the left IM. We start by the explicit representation for the left fixed point of the light-cone ($\bra{L_{\rm LC}}$) and standard ($\bra{L}$) transfer matrices in terms of the two-site gates, reading (for $t=3$)
\begin{subequations}
\begin{align}\label{eq:sketches_fp}
	\bra{L_{\rm LC}}&=\includescaledgraphics{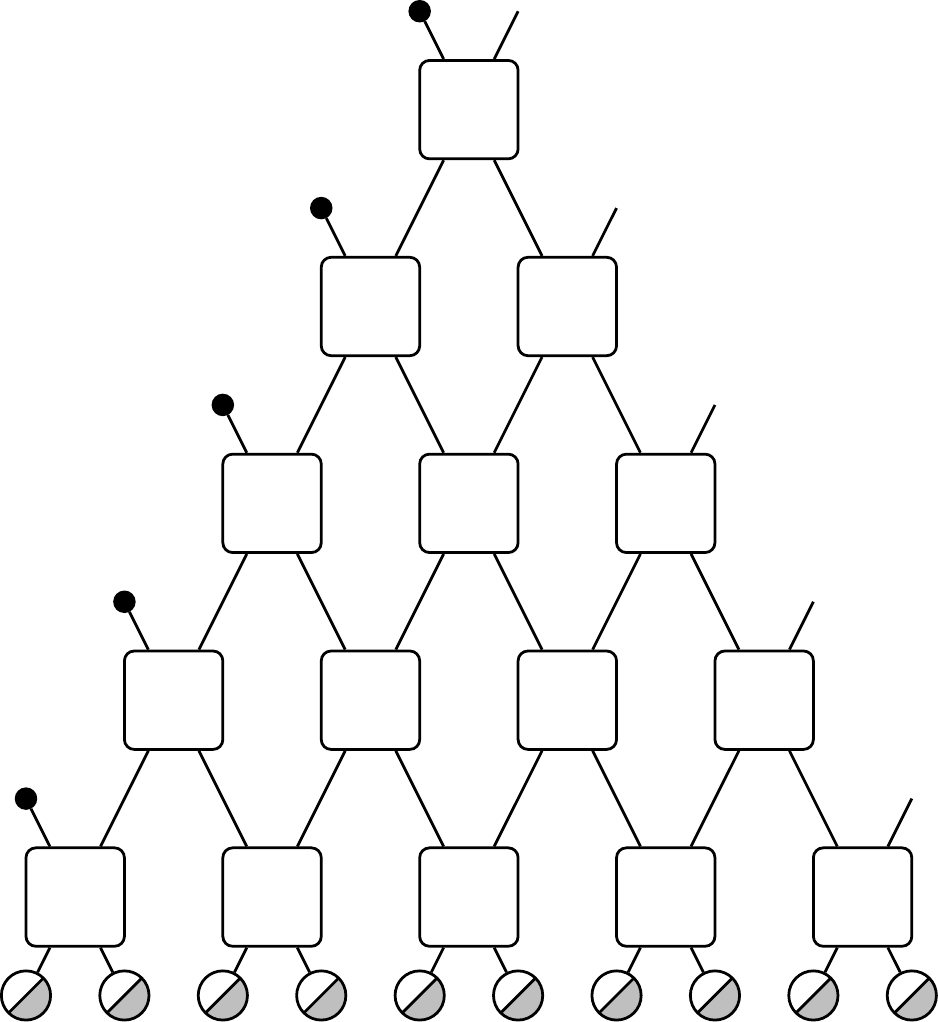} \\
	\bra{L}&=\includescaledgraphics{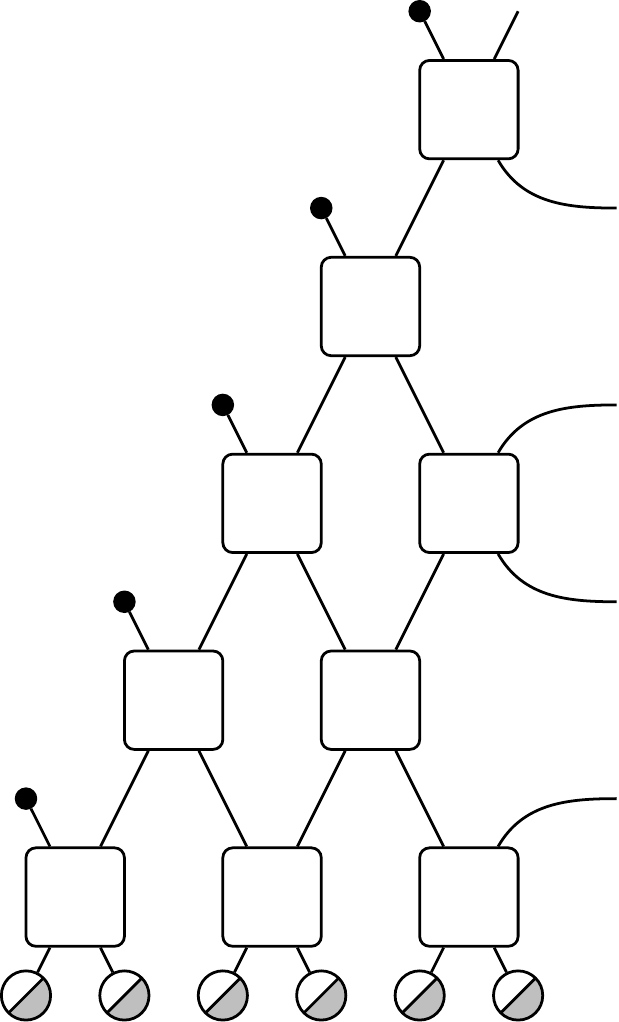}
\end{align}
\end{subequations}
We recognize that the bottom-left corner of $\bra{L}$ is the fixed point of the light-cone transfer matrix for shorter time. Therefore, using our previous result, we have
\begin{align}\label{eq:partial_result}
	\bra{L}=\includescaledgraphics{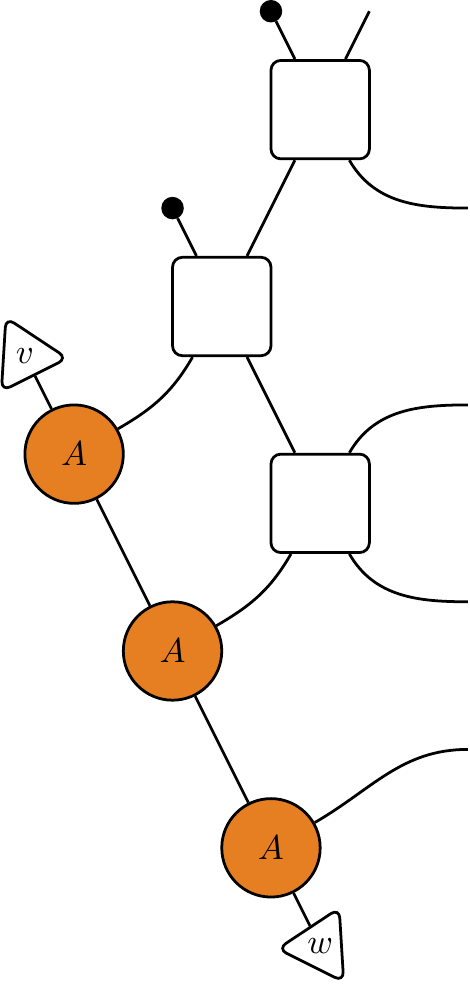}\,.
\end{align}

It is straightforward to see that the tensor $B$ satisfies the identity
\begin{align}\label{eq:id}
	\includescaledgraphics{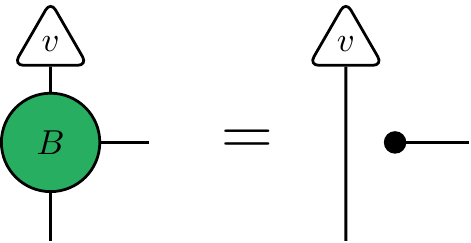}\,.
\end{align}
Applying it to Eq.~\eqref{eq:intertwinings}, this yields
\begin{align}\label{eq:id_2}
	\includescaledgraphics{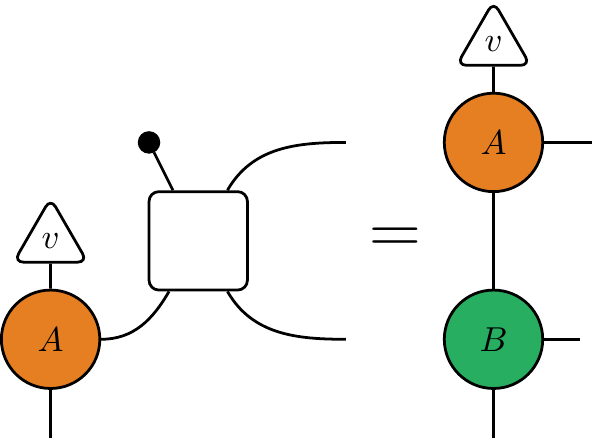}\,.
\end{align}
Finally, we can use~\eqref{eq:id_2} to simplify Eq.~\eqref{eq:partial_result}: Starting from the leftmost corner, and applying iteratively~\eqref{eq:id_2} and the zipper condition~\eqref{eq:intertwinings}, we get
\be\label{eq:final_solution}
\bra{L}= \includescaledgraphics{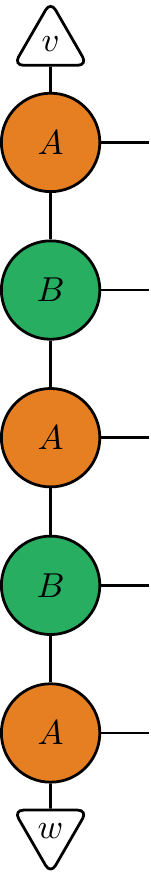}\, = \braket{v| \underbrace{A B A B \ldots B A}_{2t-1} |w}\,.
\ee
Now, because the only non-zero matrices $B_k$ are diagonal, it is immediate to see that the infinite matrices $A_k$ and $B_{k}$ can be truncated to be square matrices with $\alpha,\beta=-t,-t+1,\ldots t-1, t$.

Eq.~\eqref{eq:final_solution} holds for arbitrary values of $K$, yielding an MPS solution whose bond dimension increases linearly with $t$. However, it can be compressed to an MPS with finite bond dimension when $K/\pi$ is a rational number
\be
K/\pi=n/m\,,
\ee
with $n,m\in \mathbb{Z}$. In this case, because of the periodicity of the trigonometric and exponential functions in the tensors $A$ and $B$, it is easy to show that the corresponding matrices can be truncated to the first $m$ lines and columns. After reshaping, we obtain
\begin{subequations}
\label{eq:truncated_A}
\begin{align}
	[A_{00}]_{\alpha,\beta}&=\delta_{\alpha,\beta} \cos[2K(\alpha- 1)]\,,\\
	[A_{11}]_{\alpha,\beta}&=[A_{00}]_{\alpha,\beta}\,,\\
	[A_{01}]_{\alpha,\beta}&=\delta_{1,{\rm mod}(\alpha-\beta,m)} \cos[2K(\beta-1)]\,,\\
	[A_{10}]_{\alpha,\beta}&=\delta_{1,{\rm mod}(\beta-\alpha,m)} \cos[2K\alpha]\,,
\end{align}
\end{subequations}
and
\begin{subequations}
\label{eq:truncated_B}
\begin{align}
	[B_{00}]_{\alpha,\beta}&=\delta_{\alpha,\beta} \exp[2Ki(\alpha- 1)]\,,\\
	[B_{11}]_{\alpha,\beta}&=\delta_{\alpha,\beta} \exp[-2Ki(\alpha- 1)]\,,\\
	[B_{01}]_{\alpha,\beta}&=[B_{10}]_{\alpha,\beta}=0\,,
\end{align}
\end{subequations}
with $1\leq \alpha,\beta \leq m$.

Here we have illustrated the derivation for the left IM. A similar computation can be carried out for the right one, starting from the SW-NE light-cone transfer matrix, yielding
\be
	\ket{R} = \braket{v| \underbrace{\overline{B A B A \ldots A B}}_{2t-1} |w}\,.
\ee

\begin{figure}
	\includegraphics[width=\columnwidth]{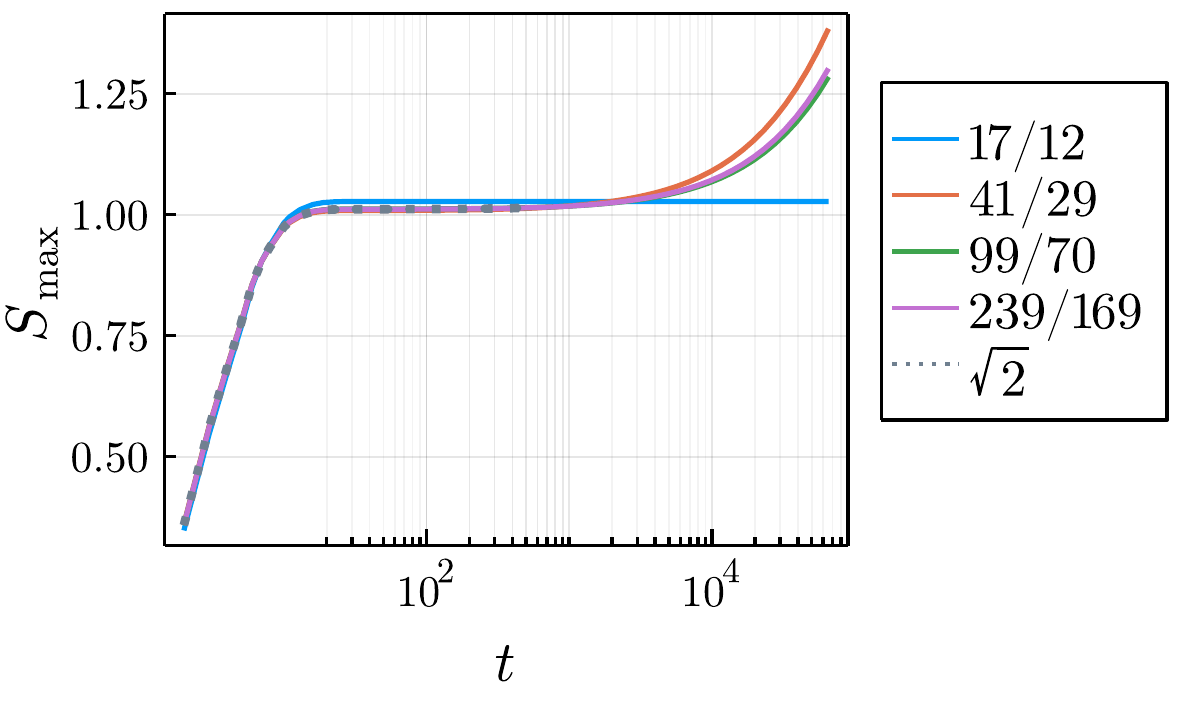}
	\caption{TE at the dual-unitary point for different values of  $K/\pi$ close to $\sqrt{2}$, quenching from $\ket{\Psi_0}=\ket{+}^{\otimes L}$. The solid lines are obtained with Eqs.~\eqref{eq:truncated_A}, \eqref{eq:truncated_B}.}
	\label{fig:exact_sm}
\end{figure}

In order to evaluate the entanglement entropy of the exact solution in Eq.~\eqref{eq:final_solution}, we resort to numerical calculations with MPS of finite size.
While the times accessible with this method are of the order of $10^5$, we notice that there can be very large finite-size effects that make the extrapolation of the asymptotic behavior sometimes challenging.
As shown in Fig.~\ref{fig:exact_sm}, plateaux of arbitrarily long times can occur when an irrational $K$ is close to a rational approximation with small denominator.
Indeed, if an irrational $K/\pi$ is $\epsilon$-close to a rational number $K_\mathrm{rat}/\pi$ with small denominator, the TE for $K/\pi$ will approximately follow the behavior of $K_\mathrm{rat}/\pi$ until a time which increases with $1/\epsilon$.
If $\epsilon$ is sufficiently small, the curve for $K_\mathrm{rat}$ may well have saturated before that time, so that the curve for $K$ will display a long initial plateau. An example of this behavior is shown in Fig.~\ref{fig:exact_sm} for $K/\pi=\sqrt{2}$. In this case, despite an initial plateau, we clearly see an eventual growth for the TE for increasingly better rational approximations of $K$. 

This example shows that, because of the very fine-tuned nature of the dual-unitary point, the behavior of the TE is extremely irregular in this case. Therefore, our numerical evidence at the dual unitary point should be taken \emph{cum grano salis}. Still, for the accessible time scales, our data consistently point to an indefinite growth of the TE for generic values of $J$.

\end{document}